\documentclass[prb,aps,twocolumn,showpacs,superscriptaddress]{revtex4}

\usepackage{amsmath,amssymb,amsthm}
\usepackage{graphicx}
\usepackage{subfigure}
\usepackage{amsmath}
\usepackage{amsfonts,amssymb}
\usepackage{bm}
\usepackage{float}
\usepackage{color}
\usepackage{sidecap}
\allowdisplaybreaks
\usepackage{soul}
\setstcolor{red}
\usepackage[normalem]{ulem}
\usepackage{hyperref}
\hypersetup{colorlinks=true,breaklinks,urlcolor=blue,linkcolor=blue,citecolor=blue}

\begin{document}

\title{Generalized Aubry-Andr\'{e} self-duality and Mobility edges in non-Hermitian quasi-periodic lattices}
\author{Tong Liu}
\thanks{t6tong@njupt.edu.cn}
\affiliation{Department of Applied Physics, School of Science, Nanjing University of Posts and Telecommunications, Nanjing 210003, China}
\author{Hao Guo}
\thanks{guohao.ph@seu.edu.cn}
\affiliation{Department of Physics, Southeast University, Nanjing 211189, China}
\author{Yong Pu}
\thanks{puyong@njupt.edu.cn}
\affiliation{Department of Applied Physics, School of Science, Nanjing University of Posts and Telecommunications, Nanjing 210003, China}
\author{Stefano Longhi}
\thanks{stefano.longhi@polimi.it}
\affiliation{Dipartimento di Fisica, Politecnico di Milano, Piazza L. da Vinci 32, I-20133 Milano, Italy}
\affiliation{IFISC (UIB-CSIC), Instituto de Fisica Interdisciplinar y Sistemas Complejos - Palma de Mallorca, Spain}

\date{\today}

\begin{abstract}
We demonstrate the existence of generalized Aubry-Andr\'{e} self-duality in a class of non-Hermitian quasi-periodic lattices with complex potentials. From the self-duality relations, the analytical expression of mobility edges is derived. Compared to Hermitian systems,  mobility edges in non-Hermitian ones not only separate localized from extended states, but also indicate the coexistence of complex and real eigenenergies, making it possible a topological characterization of mobility edges. An experimental scheme, based on optical pulse propagation in synthetic photonic mesh lattices, is suggested to implement a non-Hermitian quasi-crystal displaying mobility edges.
\end{abstract}

\maketitle

\section{Introduction} Anderson localization~\cite{Anderson}, i.e. the absence of diffusion of quantum or classical waves in disordered systems, is a milestone in condensed matter physics and beyond \cite{1a,1b,1c}.
According to the scaling theory~\cite{scaling}, in one-dimensional (1D) and two-dimensional systems with random disorder all eigenstates are exponentially localized, no matter how small the strength of disorder is, while mobility edges, separating localized and extended states in energy spectrum, are observed in three dimensional (3D) systems ~\cite{Mott}.
However, it is well known that lattices with correlated disorder can undergo a metal-insulator transition even in 1D (see, e.g., \cite{AA,2a,2b,2c,2d,2e,R1,R2,R3,R4,R5,R6,Rossignolo,Liu,Sarma,Flash,Lixp,Lix,boo} and references therein).
A paradigmatic example is provided by the famous Aubry-Andr\'{e} model~\cite{AA}, where the localization-delocalization transition can be derived from a symmetry (self-duality) argument \cite{2b}.
A hallmark of this model is the sharp nature of the
localization transition and the absence of mobility edges, i.e. all single-particle
eigenstates in the spectrum suddenly become exponentially
localized above a threshold level of disorder. Recent works reported on
mobility edges in certain quasi-periodic 1D lattices displaying a {\em generalized} Aubry-Andr\'{e} self-duality ~\cite{biddle,Ganeshan},
making it possible the observation of mobility edges in 1D systems \cite{3b,3c,3d} without resorting to 3D models.  The role of particle interaction and many-body localization in such systems have been
investigated as well \cite{4a,4b,4c,4d,4e}.  However, such previous studies have been limited to consider Hermitian models.

Non-Hermitian lattices show exotic physical phenomena without any Hermitian counterparts, such as exceptional points, breakdown of bulk-boundary correspondence based on Bloch band invariants, and non-Hermitian skin effect~\cite{Gong,Zhu,Esaki,Lee,Leykam,Shen,Yin,Yao1,Yao2,Alvarez,Kunst,Kawabata,Guo,Takata,Bender,Yuce1,Yuce2,Yuce3,Rudner,uff1,uff2,uff3,uff4,uff5}. Remarkably, disorder can behave differently in Hermitian versus non-Hermitian systems (see e.g. \cite{cazz1,cazz2,Hatano1,cazz3,cazz4} and references therein).  A seminal work dealing with disorder in non-Hermitian lattices is the Hatano-Nelson model~\cite{Hatano1,Hatano2,Hatano3}, in which
an asymmetric hopping caused by an imaginary gauge field results in a localization-delocalization transition and the existence of mobility edges \cite{Hatano4}.
Since this pioneering study, several non-Hermitian models with either random or incommensurate disorder have been investigated,
in which non-Hermiticity is introduced by either asymmetric hopping amplitudes or complex on-site potentials~\cite{Jazaeri,Yuce,Liang,Jiang,Liuy,Zeng1,Zeng2,Zeng3,Wangr,Longhi1,Longhi2,new2}. In certain models, the topological nature of the localization transition and self-duality have been discussed \cite{Gong,Longhi1,new2}.
{However, we emphasize that the self-dual symmetry falls into two categories.
The first category is the {\em generalized} Aubry-Andr\'{e} self-dual symmetry. Models possessing such a generalized symmetry show mobility edges, and their analytical form can be obtained from the self-dual relations. Such models are rare and there are only few know examples for Hermitian systems ~\cite{biddle,Ganeshan}. The second category is the ``simple" self-dual symmetry. This is a much more common kind of symmetry which is found in many models~\cite{new2,MKohmoto,Sun,Gopalakrishnan}, including some non-Hermitian systems. However, this type of symmetry cannot be used to derive analytical form of mobility edges, and so far available non-Hermitian models displaying mobility edges \cite{Hatano1,Hatano4,Gong,new2} resort to numerical results.
Here a major question arises: can a {\em generalized} Aubry-Andr\'{e} self-dual symmetry and exact form of mobility edges be found beyond Hermitian quasi-crystals?\\
\\
In this work we address this open question and introduce exactly-solvable non-Hermitian models in quasi-periodic lattices with complex potentials displaying mobility edges and a generalized Aubry-Andr\'{e} self-duality. A photonic implementation of the proposed models, based on pulse propagation in synthetic fiber mesh lattices, is also presented.

\section{Exactly-solvable non-Hermitian models displaying generalized Aubry-Andr\'{e} self-duality}
\subsection{Model I}
\begin{figure}
  \centering
  \includegraphics[width=0.5\textwidth]{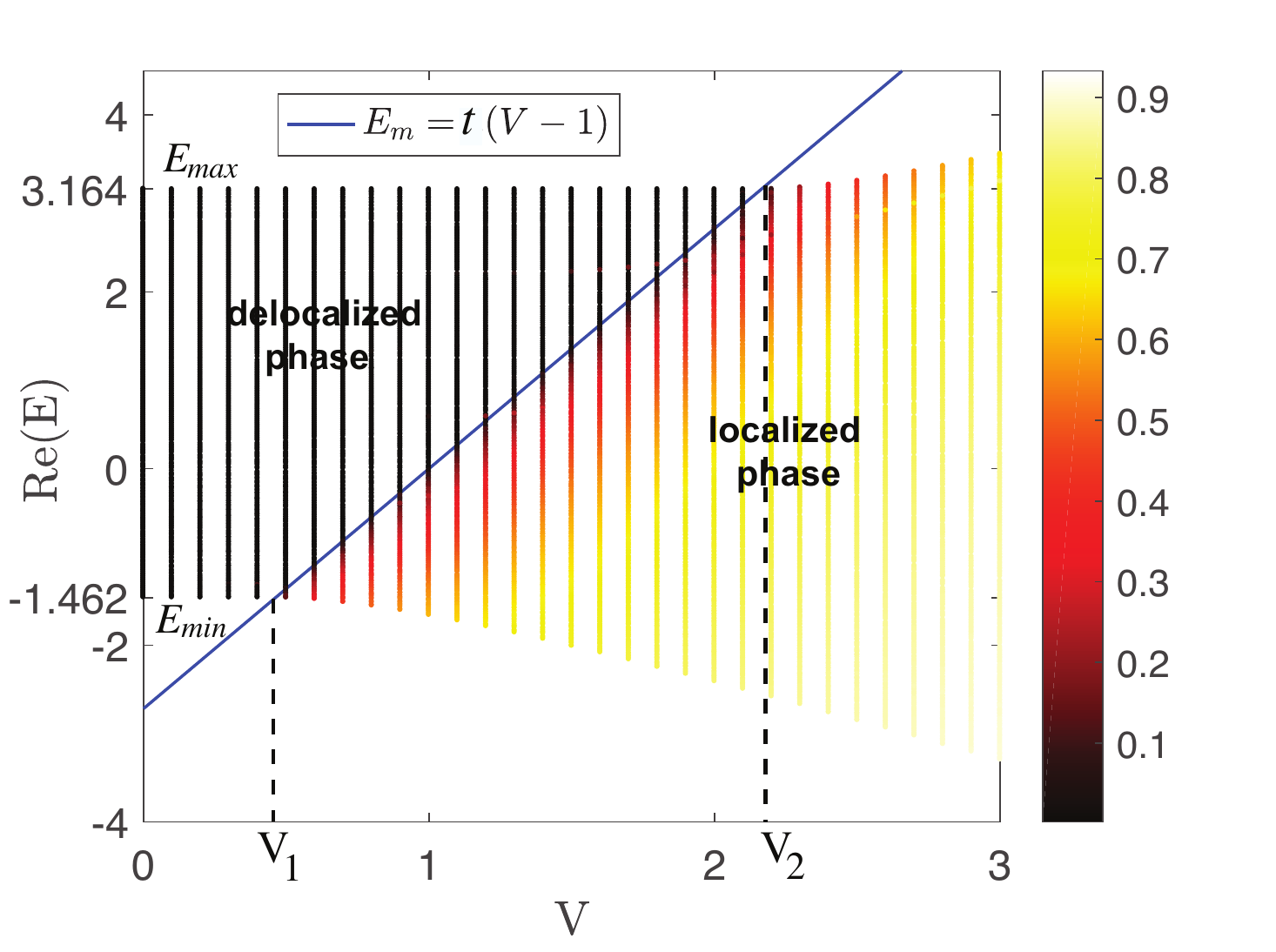}\\
  \caption{(Color online) The real part of eigenvalues of Eq.~(\ref{eq1}) and IPR as a function of $V$ with the parameter $s=1$. The total number of sites is set to be $L=500$. Different colours of the eigenvalue curves indicate different magnitudes of the IPR of the corresponding wave functions. The black eigenvalue curves denote the delocalized states, and the bright yellow eigenvalue curves denote the localized states. The blue solid lines represent the boundary between spatially localized and delocalized states, i.e., the mobility edge $E_m=t(V-1)$. In the delocalized phase, the upper ($E_{max}$) and lower  ($E_{min}$) boundaries of the energy spectrum are independent of $V$.
  }
  \label{001}
\end{figure}
As a first example of generalized Aubry-Andr\'{e} self-duality, let us consider a non-Hermitian 1D model with exponentially-decaying hopping amplitude and quasi-periodic complex on-site potential, defined by the eigenvalue equation
\begin{equation}
(E+t) \psi_n = t \sum_{n'}e^{-s\mid n-n'\mid}\psi_{n'}+V e^{i 2\pi\alpha n}\psi_n,
\label{eq1}
\end{equation}
where $s>0$ is the decay rate of the non-nearest-neighbor hopping, $V$ is the complex potential strength, $\alpha$ is irrational, and $\psi_n$ is the amplitude of wave function at the $n$th lattice. We choose the parameters $\alpha=(\sqrt{5}-1)/2$ and $t=e^{s}$.
When the on-site potential is replaced by $V\cos(2\pi\alpha n)$, this model becomes the Hermitian quasi-periodic lattice studied in Ref.~\cite{biddle}, which displays mobility edges given by $E_m=\cosh(s)V - e^{s}$. To determine the expression of mobility edges in the non-Hermitian case, we first introduce a function $W_n(s_0)$ defined as
\begin{equation}
\begin{aligned}
&E+t-Ve^{i 2\pi\alpha n} = (E+t)W_n,\\
&e^{s_0}=\frac{E+t}{V},~W_n(s_0)=\frac{e^{s_0}-e^{i 2\pi\alpha n}}{e^{s_0}},
\label{eq2}
\end{aligned}
\end{equation}
so that Eq.~(\ref{eq1}) takes the form
\begin{equation}
(E+t)W_n(s_0) \psi_n = t \sum_{n'} e^{-s\mid n-n'\mid}\psi_{n'}.
\label{eq3}
\end{equation}
After multiplying both sides of Eq. (\ref{eq3}) by $W_k(s)e^{i k 2\pi\alpha n}$, summing over $n$ and setting $\phi_k = \sum_{n} e^{ 2\pi i \alpha k n}W_n(s_0) \psi_n$,
one obtains
\begin{equation}
(E+t)W_k(s) \phi_k = t \sum_{k'}e^{-s_0\mid k-k'\mid}\phi_{k'}.
\label{eq7}
\end{equation}
The detailed derivation of Eq. (\ref{eq7}) is given in Appendix A. Remarkably, when $s=s_0$, Eq.~(\ref{eq1}) has the same form as Eq.~(\ref{eq7}), i.e. a generalized  Aubry-Andr\'{e}  self-duality is found. Following the Aubry-Andr\'{e} work \cite{AA}, we conjecture that the localization-delocalization transition is located at the self-dual point $\frac{E+t}{V}=e^{s}$.
Thus the non-Hermitian quasi-periodic lattice defined by Eq.~(\ref{eq1}) displays a mobility edge at the energy
\begin{equation}
E_m=t(V-1).
\label{eq8}
\end{equation}
To verify our conjecture, we calculated analytically the Lyapunov exponent $\mu(E)$ for the eigenstates of Eq. (\ref{eq1}), and found that the mobility edge obtained from Lyapunov exponent analysis is precisely given by Eq. (\ref{eq8}); technical details are given in Appendix B.  Remarkably, the self-duality argument enables to analytically compute mobility edges without solving the spectral problem. The following bounds $E_{min}<E<E_{max}$  for the energy spectrum $E$ of the delocalized phase can be derived (Appendix B), where
\begin{equation}
\label{eq9}
E_{min}=-\frac{2 t}{t+1} \; , \;\; E_{max } =\frac{2t}{t-1},
\end{equation}
indicating that the delocalized modes correspond to real energies. Conversely, for the localized modes the energy spectrum is complex. In other words, the mobility edge $E_m$ not only discriminates between localized and delocalized states, but also between real and complex energies. A major implication of this result is that a  {\em {topological}} number can be introduced to predict the existence of a mobility edge. This entails to compute winding numbers $w(E_B)$ that count the number of times the complex spectral trajectory encircles a base energy $E_B$ as an external phase in the Hamiltonian is varied  \cite{Gong,Longhi1,new2}. As shown in Appendix C, the knowledge of the winding numbers at the two base energies $E_B=E_{min}$ and $E_B=E_{max}$ is sufficient to topologically predict the existence of the mobility edge.

Our theoretical predictions have been verified by a numerical analysis of Eq. (\ref{eq1}) on a finite lattice containing $L$ sites under periodic boundary conditions. The localization properties of eigenstates are measured by the inverse participation ratio (IPR)~\cite{IPR}. For a normalized wave function, it is defined as
\[
\text{IPR}_n =\sum_{j=1}^{L} \left|\psi^n_{j}\right|^{4}, \] where $n$ is the index of energy level.
It is well known that the IPR of a delocalized state scales like $L^{-1}$, thus vanishing in the thermodynamic limit, while it is finite for a localized state. In Fig.~\ref{001} we show the numerically-computed  IPR diagram in the $({\rm Re}(E),V)$ plane on a pseudo color map, clearly demonstrating a mobility edge along the line defined by Eq. (\ref{eq8}). The mobility edge also separates real and complex energies.

\subsection{Model II}
\begin{figure}
  \centering
  \includegraphics[width=0.5\textwidth]{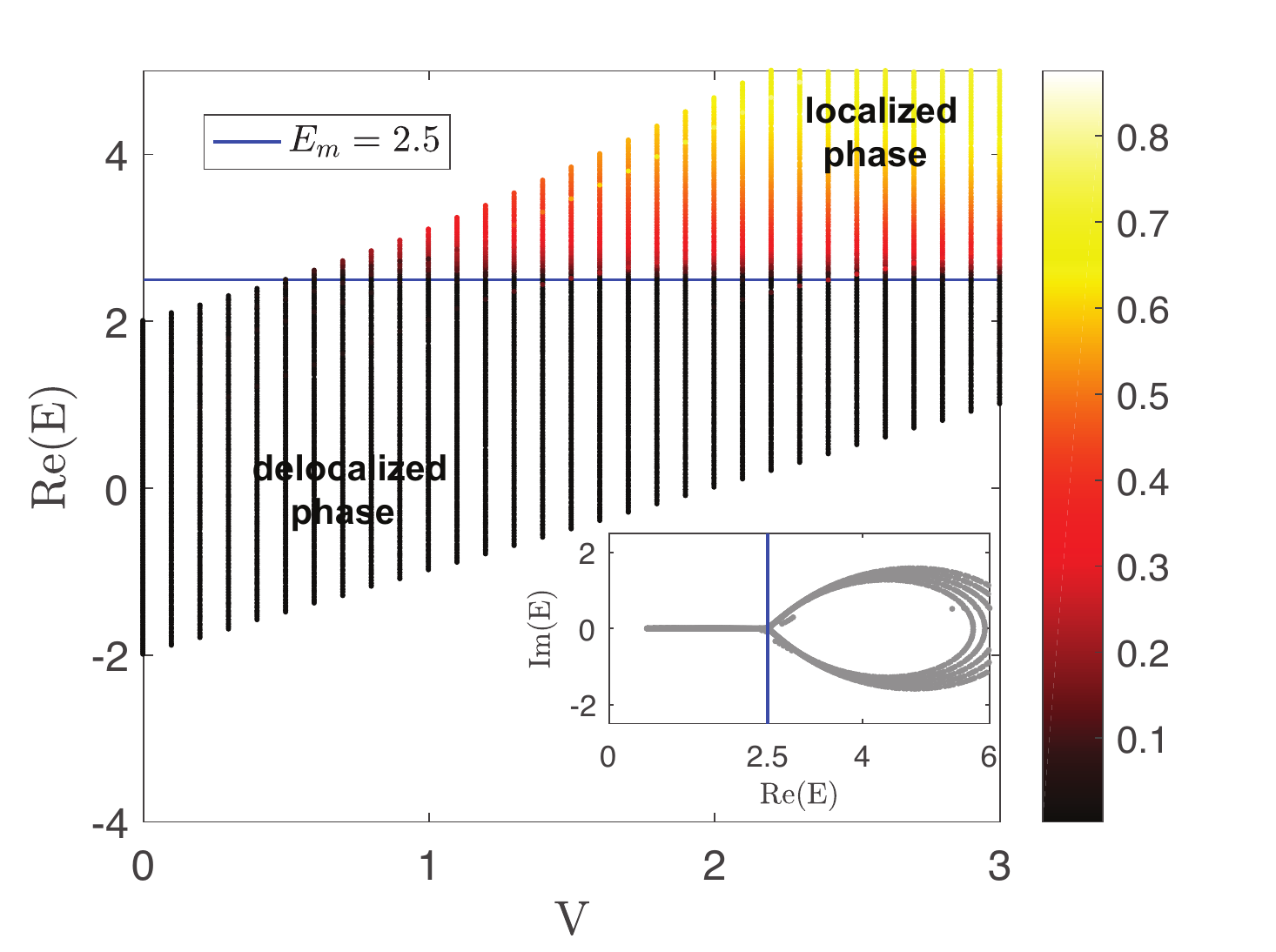}\\
  \caption{(Color online) The real part of eigenvalues of Eq.~(\ref{Eq1}) and IPR as a function of $V$ with the tuning parameter $a=0.5$. The total number of sites is set to be $L=500$. Different colours of the eigenvalue curves indicate different magnitudes of the IPR of the corresponding wave functions. The black eigenvalue curves denote the delocalized states, and the bright yellow eigenvalue curves denote the localized states. The blue solid lines represent the boundary between spatially localized and delocalized states, i.e., the mobility edge $E_m=a+1/a$. In the inset, it is clearly shown that the mobility edge $E_m=2.5$ separates the real and complex energy spectrum.}
  \label{002}
\end{figure}
As a second example, let us consider a nearest-neighbor hopping model with tunable complex on-site potential defined by the eigenvalue equation
\begin{equation}
E \psi_n=\psi_{n+1}+\psi_{n-1}+\frac{V}{1-a e^{i 2\pi\alpha n}} \psi_n,
\label{Eq1}
\end{equation}
where  $0<a<1$ is a potential tuning parameter; other parameters are the same
as in Eq.~(\ref{eq1}). This model provides a non-Hermitian extension of the quasi-periodic lattice previously introduced in Ref.~\cite{Ganeshan} and displaying generalized self-duality. As compared to model I, model II is more feasible for an experimental implementation since it does not require
 hopping control, but it is not amenable for a full analytical treatment  (Lypaunov exponent calculation). As we are going to show, model II displays generalized self-duality, which
 is enough to analytically predict mobility edges. To this aim,
let us multiply both sides of Eq.~(\ref{Eq1}) by $e^{i 2\pi\alpha n m}$ and sum over $n$. After setting $\phi_m=\sum_n e^{i 2\pi\alpha n m}\psi_n$, from Eq.~(\ref{Eq1}) one obtains
\begin{equation}
\label{Eq2}
[E-2 \cos(2\pi\alpha m)]\phi_m=\sum_n e^{i 2\pi\alpha n m} \frac{V}{1-a e^{i 2\pi\alpha n}} \psi_n.
\end{equation}
We introduce some functions defined as follows
\begin{equation}
\begin{aligned}
\label{Eq3}
&e^{s_0} = \frac{1}{a},~W_n(s_0)=\frac{e^{s_0}-e^{i 2\pi\alpha n}}{e^{s_0}},\\
&E =  2\cosh(s),~\Omega_m(s)=\frac{\cosh(s)-\cos(2\pi\alpha m)}{\sinh(s)},
\end{aligned}
\end{equation}
so that Eq.~(\ref{Eq2}) can be written as
\begin{equation}
\label{Eq4}
2\sinh(s)\Omega_m(s)\phi_m=V \sum_{r'} e^{-\mid m-r'\mid s} \phi_{r'}.
\end{equation}
Multiplying both sides of Eq.~(\ref{Eq4}) by $\Omega_k(s_0)e^{i 2\pi\alpha m k}$, summing over $m$ and after setting $\varphi_k=\sum_m e^{i 2\pi\alpha m k}\Omega_m(s)\phi_m$, Eq.~(\ref{Eq4}) takes the form
\begin{equation}
\label{Eq5}
2\sinh(s)\Omega_k(s_0)\varphi_k= V\sum_{k'} e^{-\mid k-k'\mid s}\varphi_{k'}.
\end{equation}
Finally, let us  multiply both sides of Eq. (\ref{Eq5}) by $e^{i 2\pi\alpha k q}$, summing over $k$ and setting $\mu_q=\sum_k e^{i 2\pi\alpha k q} \varphi_k$, Eq. ~(\ref{Eq5}) can be transformed as
\begin{equation}
\label{Eq6}
2\frac{\sinh(s)}{\sinh(s_0)}\cosh(s_0)\mu_q=\mu_{q-1}+\mu_{q+1}+V \frac{e^{s}}{e^{s}-e^{i 2\pi\alpha q}} \mu_q.\\
\end{equation}
Note that when $s=s_0$, Eq. (\ref{Eq1}) has the same form as Eq. (\ref{Eq6}). From the self-dual relations $e^{s}=1/a$ and $E=2 \cosh(s)$, we obtain the mobility edge energy
\begin{equation}
E_m=a+1/a .
\label{Eq7}
\end{equation}
Interestingly, for a fixed value of the tuning parameter $a$, $E_m$ is independent of the potential strength $V$. The property of ``being constant" of the mobility edge is a remarkable result, not reported in the literature yet. We checked the predictions of the theoretical analysis by direct numerical simulations of Eq.~(\ref{Eq1}). The numerical results, shown in  Fig.~\ref{002}, confirm our theoretical predictions with excellent accuracy. From the analysis of the energy spectrum, we find the same scenario like model I, i.e. extended (localized) eigenstates correspond to real (complex) energies (see the inset of Fig.~\ref{002}). A winding number, revealing the topological signature of the mobility edge, can be also introduced for this model as well, as shown in Appendix C.

\begin{figure}
  \centering
  \includegraphics[width=0.4\textwidth]{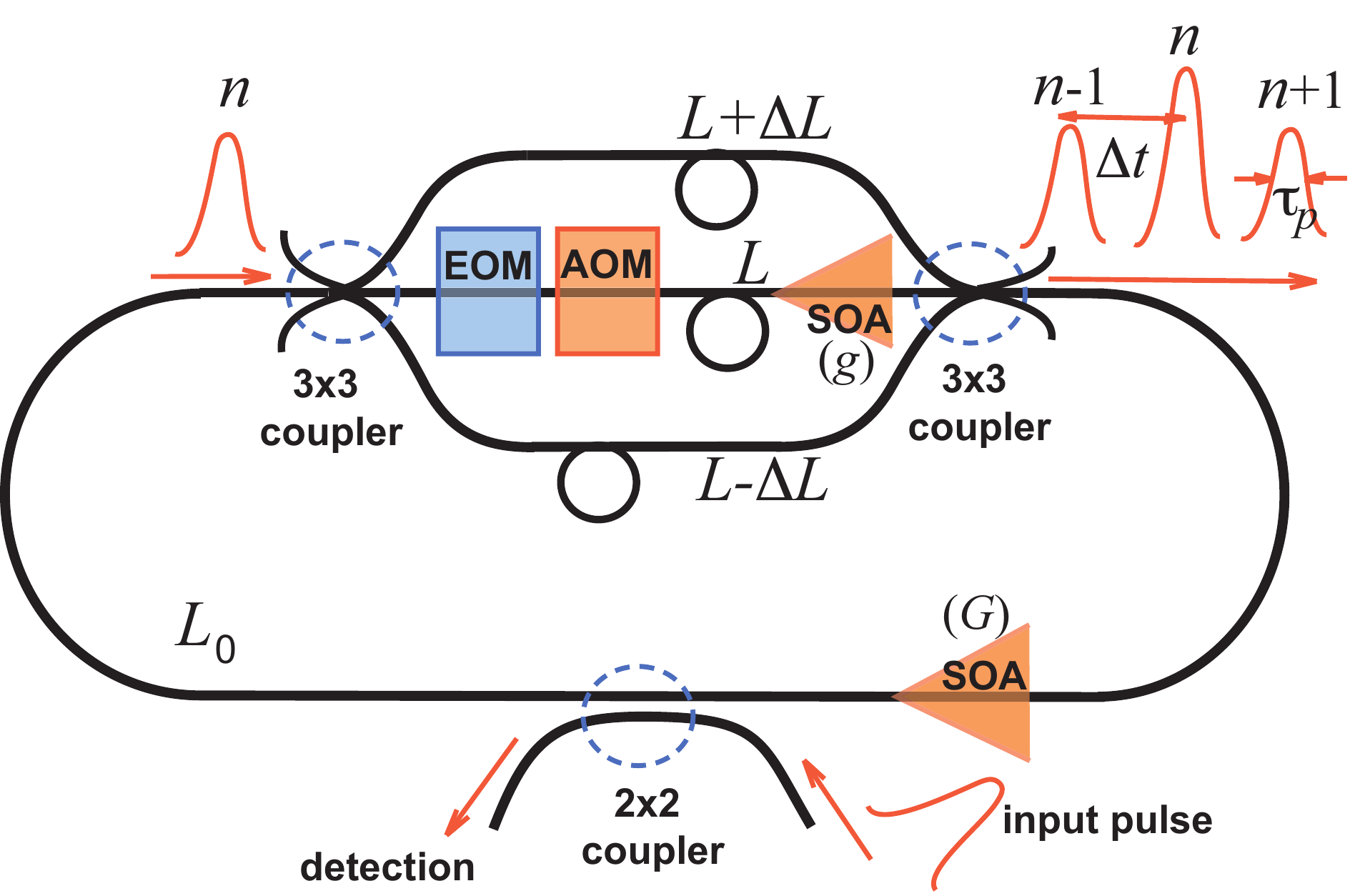}\\
  \caption{(Color online) Schematic of  a synthetic mesh photonic lattice with complex potential, based on pulse propagation in fiber loops. An optical pulse propagating in the main fiber loop of length $L_0$ is splitted into three pulses, after each transit, via a three-arm fiber interferometer with unbalanced arms of length $L$ and $L \pm \Delta L$, with $L \ll L_0$. Time pulse separation $\Delta t= \Delta L /c$, introduced by arm length unbalance, provides the time slot of the lattice mesh. The interferometer is coupled to the main fiber loop via two $3 \times 3$ symmetric fiber couplers. A non-dispersive optical pulse of duration $\tau_p< \Delta t /2$ is initially injected into the loop via a $2 \times 2$ fiber coupler. The dynamical evolution of the pulse amplitudes $a^{(m)}_n$ at successive transits $m$ in the loop can be monitored by the output port of the coupler. The synthetic complex potential is obtained by placing amplitude (AOM) and phase (EOM) modulators in the central arm of the interferometer and driven by independent step-wise waveforms $h_n^{(AOM)}$ and $h_n^{(EOM)}$.
Two semiconductor optical amplifiers (SOA), with  gain parameters $g$ and $G$, are also included in the central arm of the interferometer and in the main loop.}
  \label{FigS1}
\end{figure}

\section{Proposal of experimental implementation}
Photonic systems have been recently shown to provide an experimentally-accessible platform to implement non-Hermitian lattices and to observe a wide variety of non-Hermitian phenomena, such as parity-time symmetry breaking, exceptional points, non-Hermitian skin effect etc.
To observe mobility edges in non-Hermitian quasi-periodic potentials, we focus our attention to model II discussed in previous section, which is more amenable for an experimental realization. We consider discrete-time quantum walk of optical pulses in synthetic photonic mesh lattices \cite{RS1,RS2,RS3,RS4,RS5,RS6,RS7,RS8},  realized in coupled fiber rings with unbalanced path-lengths. Such synthetic lattices have been
experimentally used to demonstrated a wealth of phenomena, such as Bloch oscillations \cite{RS3}, parity-time symmetric phase transitions \cite{RS4,RS5}, Anderson localization \cite{RS2,RS6,RS7}, and the  non-Hermitian skin effect \cite{RS8}.  By proper combination of amplitude and phase modulators in the fiber loops \cite{RS4,RS7,RS8},  they can engineer rather arbitrary non-Hermitian potentials. A schematic of the synthetic photonic lattice is shown in Fig \ref{FigS1}.  A short pulse is
launched, via a $2 \times 2$ optical coupler, into a single mode fiber loop of length $L_0$, which is coupled to a three-arm fiber interferometer of mismatched lengths $L$ (central arm)  and $L \pm \Delta L$ (upper and lower arms) by  $3 \times 3$
symmetric fiber couplers, with  $L \ll L_0$. The central arm of the interferometer includes a semiconductor optical amplifier (SOA), an electro-optic phase modulator (EOM) and an acousto-optic amplitude modulator (AOM).  A second SOA is also placed in the main loop of length $L_0$. At each transit in the main loop, a pulse  entering into the interferometer  is splitted, at the output port, into three pulses with time delays $-\Delta t$, $0$ and $\Delta t$, where $\Delta t=\Delta L/c$ is the time delay introduced by the unbalanced arms in the interferometer. The successive pulse splitting emulates a discrete-time quantum walk, where the complex amplitude $a_n^{(m)}$ of pulse occupying the $n$-th time slot (discrete space distance) at the $m$-th round trip evolves according a linear map \cite{RS1,RS2,RS3,RS4,RS5,RS6,RS7}.
In particular, by tailoring the EOM and AOM signals, one can implement a non-Hermitian Hamiltonian with nearest-neighbor hopping and a rather arbitrary complex on-site potential $V_n$, such as the quasi-periodic potential in Eq. (\ref{Eq1}). Details are given in Appendix D. Pulse evolution measurements at successive transits in the loop, detected by the output port of the $2 \times 2$ optical coupler (Fig. \ref{FigS1}), can provide a clear signature of the presence (or absence) of the mobility edge. If all eigenstates of the system are delocalized and the corresponding energy spectrum entirely real, an initial pulse will spread across the synthetic lattice at successive transits. This scenario is illustrated in Fig.~\ref{FigS3}(a). The figure depicts on a pseudo color map the numerically-computed evolution, at successive discrete time steps $m$, of the normalized pulse amplitudes $|a^{(m)}| / \sqrt{P_m}$, with $P_m=\sum_n |a_n^{(m)}|^2$, as obtained from the map, defined by Eq. (D-3) of Appendix D, for the initial condition $a_n^{(0)}=\delta_{n,0}$ (a single pulse is injected into the main fiber loop) and for $V=0.4$, $a=0.5$. According to Fig.\ref{002}, for such parameter values all eigenmodes are delocalized and the energy spectrum entirely real. On the other hand, in the presence of a mobility edge some eigenstates are localized with corresponding complex energies. In this case the localized modes with the highest growth rate (i.e. imaginary part of the eigenenergy) will dominate, resulting in a frozen spreading of $|a^{(m)}|$ as the time step $m$ increases. This case is illustrated in Fig.~\ref{FigS3}(b), corresponding to $V=1$ and $a=0.5$.
\begin{figure}
  \centering
  \includegraphics[width=0.45\textwidth]{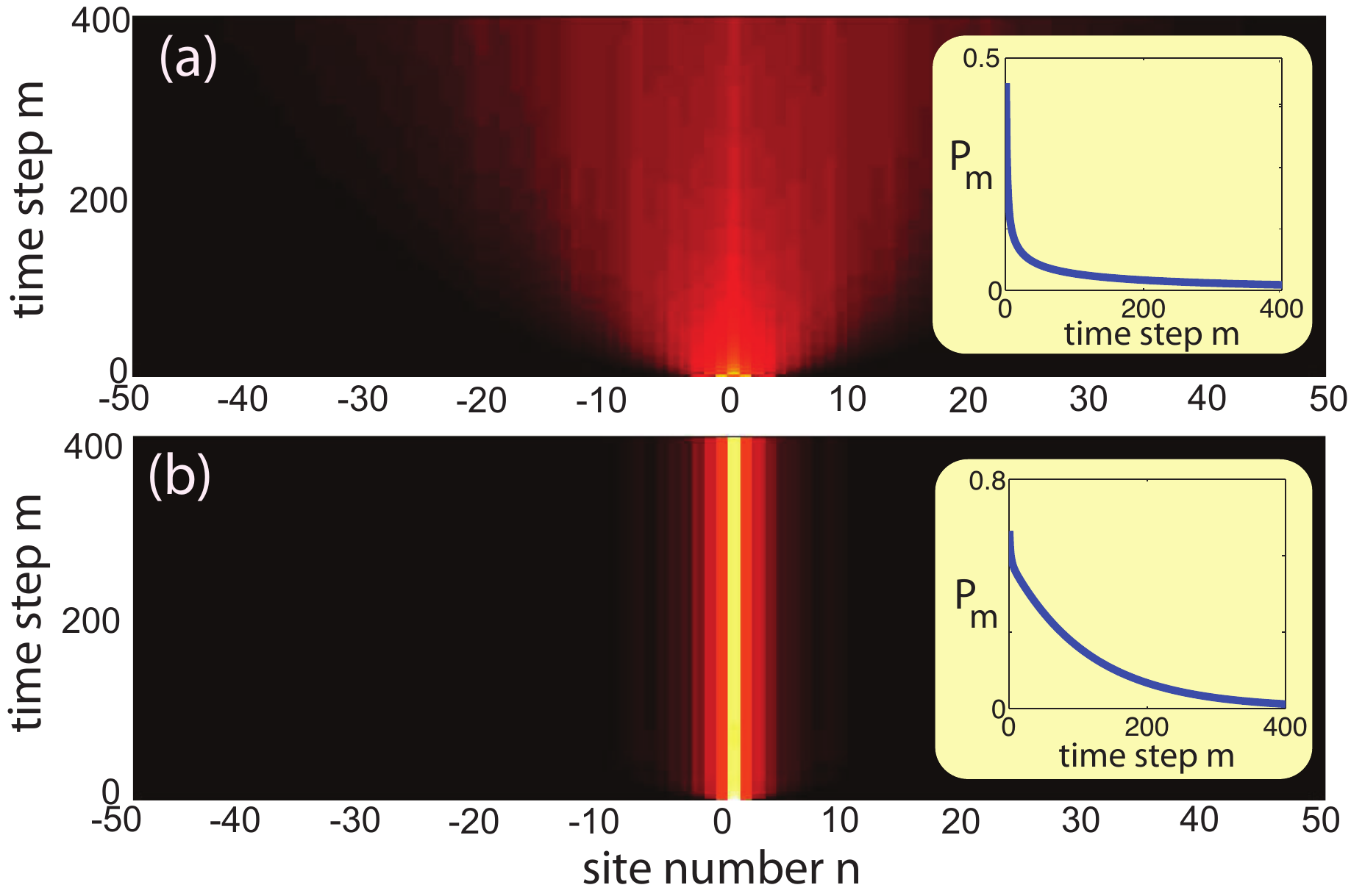}\\
  \caption{(Color online) Behavior of normalized pulse amplitudes $|a_n^{(m)}| / \sqrt{P_m}$ (with $P_m=\sum_n |a_n^{(m)}|^2$) on a pseudocolor map for the complex potential of Model II with (a)
  $V=0.2$, $a=0.5$ (all eigenmodes are delocalized), and (b) $V=1$, $a=0.5$ (there is a mobility edge). The insets show the behavior of total optical power $P_m$ versus time steps $m$ for a gain parameter $G=0.656$ in (a), and $G=0.31$ in (b).}
  \label{FigS3}
\end{figure}

\section{ Conclusions} In this work we unveiled a class of non-Hermitian quasi-periodic lattices displaying generalized Aubry-Andr\'{e} self-duality and provided for the first time the analytic form of mobility edges in any non-Hermitian disordered system. An experimental scheme to observe mobility edges, accessible with current photonic technologies, has been proposed. The self-dual symmetry has a simple structure but a profound significance in non-Hermitian models since it predicts {\em both} the boundary of critical states {\em and} the transition from real to complex energy spectrum, thus enabling to introduce a {\em topological} signature of mobility edges. Our results push the  concept of generalized self-duality beyond known Hermitian models, providing a major tool to explore the rich physics of non-Hermitian systems with correlated disorder.

\begin{acknowledgments}
This work is supported by the National Natural Science Foundation of China (Grants No. 61874060, 11674051, and U1932159), Natural Science Foundation of Jiangsu Province (Grant No. BK2020040057), and NUPTSF (Grant No. NY217118).
\end{acknowledgments}

\appendix
\section { Derivation of Eq.~(4)}
In this Appendix we provide some technical details leading to Eq.~(4) given in the main text. Multiplying  both sides of Eq.~(3) by $W_k(s)e^{i k 2\pi\alpha n}$ and summing over $n$,
one obtains
\begin{equation}
(E+t)W_k(s) \phi_k = S_{rh}
\label{S1-0}
\end{equation}
where we have set
\begin{equation}
S_{rh}=W_k(s)\sum_{n} e^{i k 2\pi\alpha n} t \sum_{n'}e^{-s\mid n-n'\mid}\psi_{n'}.
\label{S1-1}
\end{equation}
With the substitution $r=n-n'$, one obtains
\begin{equation}
S_{rh} = t\sum_{n'} e^{i k 2\pi\alpha n'} W_k(s)\sum_{r} e^{i k 2\pi\alpha r}e^{-s \mid r\mid}\psi_{n'}.
\label{S1-2}
\end{equation}
Using the identity
\begin{equation}
W_k(s)^{-1}=\frac{e^{s}}{e^{s}-e^{i 2\pi\alpha k}} = \sum_{r} e^{-\mid r\mid s} e^{i r 2\pi\alpha k},
 \label{S1-3}
\end{equation}
one has
\begin{equation}
S_{rh} = t\sum_{n'} e^{i n' 2\pi\alpha k}\psi_{n'}.
 \label{S1-4}
\end{equation}
Clearly, since $W_{n'}(s_0)^{-1} W_{n'}(s_0)=1$, and can also write
\begin{equation}
\begin{aligned}
S_{rh} &= t\sum_{n'}W_{n'}(s_0)^{-1} W_{n'}(s_0) e^{i n' 2\pi\alpha k}\psi_{n'}\\
& = t\sum_{n'}\sum_{r} e^{-\mid r\mid s_0} e^{i r 2\pi\alpha n'} W_{n'}(s_0) e^{i n' 2\pi\alpha k}\psi_{n'}\\
& = t\sum_{n'}\sum_{r} e^{-\mid r\mid s_0} e^{i (r+k) 2\pi\alpha n'} W_{n'}(s_0)\psi_{n'}.\\
\label{S1-5}
\end{aligned}
\end{equation}
Finally, after the substitution $k'=r+k$, one obtains
\begin{equation}
\begin{aligned}
S_{rh} &= t\sum_{k'} e^{-\mid k-k'\mid s_0}\sum_{n'}e^{i n'2\pi\alpha k'} W_{n'}(s_0)\psi_{n'}\\
& = t\sum_{k'} e^{-\mid k-k'\mid s_0}\phi_{k'}.
\label{S1-6}
\end{aligned}
\end{equation}
Substitution of Eq.~(\ref{S1-6}) into Eq.~(\ref{S1-0}) yields Eq.~(4) given in the main text.\\

\section{ Lyapunov exponent analysis (Model I)}
In this Appenix we provide exact analytical computation of the energy-dependent Lyapunov exponent for the solutions to Eq.~(1) with {\em real} eigenvalue $E$. To this aim, let us assume periodic boundary conditions on a ring of size $L$ with $\alpha L \sim$ integer, i.e., $\psi_{n+L}=\psi_n$, and then take the $L\rightarrow\infty$ limit. We consider the discrete Fourier transform
\begin{equation}
\begin{aligned}
\label{S2-1}
&\phi_n= \frac{1}{\sqrt{L}} \sum_{l=1}^L e^{-2\pi i \alpha l n}\psi_l,\\
&\psi_n= \frac{1}{\sqrt{L}} \sum_{l=1}^L e^{ 2\pi i \alpha l n}\phi_l,\\
\end{aligned}
\end{equation}
so that in the dual space Eq.~(1) yields the following difference equation for $\phi_n$,
\begin{equation}
E \phi_n = \Omega_{n} \phi_{n}+V \phi_{n-1},
\label{S2-2}
\end{equation}
where e have set $\Omega_{n}=t \sum_{l\neq0}e^{-\mid l\mid s} e^{ -2 \pi i \alpha l n }=2 t {\rm Re} \{\frac{\beta_n}{1-\beta_n}\}$, $\beta_n=e^{-s-2\pi i \alpha n}$. The explicit expression of $\Omega_{n}$ reads
\begin{equation}
\Omega_{n}=2 t \frac{t \cos(2\pi\alpha n)-1}{1-2t\cos(2\pi\alpha n)+t^2},
\label{S2-3}
\end{equation}
For an arbitrary integer $n_0$, a formal solution with the eigenvalue $E=\Omega_{n_0}$ of Eq.~(\ref{S2-2}) is given by
\begin{equation}
\label{S2-4}
\phi_n \propto \left\{
\begin{array}{cc}
0 & n <n_0 \\
1  & n=n_0 \\
\frac{V}{E-\Omega_n}& n >n_0
\end{array}
\right.
\end{equation}

Let us calculate the Lyapunov exponent $\mu$ of the eigenfunction~(\ref{S2-4}) in dual space with the eigenvalue $E=\Omega_{n_0}$
\begin{equation}
\label{S2-5}
\mu(E) =  - \lim_{n \rightarrow \infty} \frac{1}{n-n_0} \log \left| \frac{\phi_{n}}{\phi_{n_0}} \right|,
\end{equation}
with $\mu(E)>0$ for $\sum_n | \phi_{n} |^2<\infty$, i.e., $\mu(E)>0$ corresponds to the {\em localization} in dual space {\em and} the {\em delocalization} in real space. From Eq.~(\ref{S2-4}) and~(\ref{S2-5}), one obtains
\begin{equation}
\label{S2-6}
\mu(E) =  \lim_{n \rightarrow \infty} \frac{1}{n-n_0} \sum_{k=n_0+1}^{n} \log \left| \frac{\Omega_k-\Omega_{n_0}}{V} \right| .
\end{equation}
After setting
\begin{equation}
\label{S2-7}
F(q) \equiv  \frac{t \cos(q)-1}{1-2t\cos(q)+t^2}
\end{equation}
so that $\Omega_{k}=2 t F(q=2\pi\alpha k)$, using the Weyl$^{\prime}$s equidistribution theorem of irrational rotations one can write
\begin{equation}
\label{S2-8}
\mu(E) =  \frac{1}{2 \pi}\int_{-\pi}^{\pi} dq \log \left| 2 t \frac{F(q)-F(q_0)}{V} \right| .
\end{equation}
with $q_0=2\pi\alpha n_0$, i.e.,
\begin{equation}
\label{S2-9}
\mu(E) = \log \left( \frac{2 t}{V} \right) + \frac{1}{2\pi}\int_{-\pi}^{\pi} dq \log \left| F(q)-F(q_0)\right| .
\end{equation}
Taking into account that
\begin{equation}
\begin{aligned}
\label{S2-10}
&\frac{1}{2\pi}\int_{-\pi}^{\pi} dq \log \left| F(q)-F(q_0)\right|\\
&=\frac{1}{2\pi} {\rm Re} \left\{ \int_{-\pi}^{\pi} dq \log(\frac{t \cos(q)-1}{1-2t\cos(q)+t^2}-\sigma) \right\},
\end{aligned}
\end{equation}
with $\sigma\equiv \Omega_{n_0}/(2t)=E/(2t)$, the integral on the right side of Eq.~(\ref{S2-10}) can be computed in a closed form to give
\begin{equation}
\label{S2-11}
\frac{1}{2\pi}\int_{-\pi}^{\pi} dq \log \left| F(q)-F(q_0)\right|=\log \left( \frac{2\sigma+1}{2t} \right)
\end{equation}
so that
\begin{equation}
\label{S2-12}
\mu(E) =\log \left( \frac{  E/t+1}{V} \right).
\end{equation}
 The real energy $E$ belongs to the spectrum of the Hamiltonian, with {\em delocalized} eigenstate in real space, provided that $\mu(E)>0$. Using Eq.~(\ref{S2-12}), this condition yields $E>E_m$,
 where we have set $E_m = t(V-1)= \exp(s) (V-1)$. Remarkably, the energy boundary $E=E_m$ corresponds to the mobility edge given by Eq.~(5), derived using the self-duality argument.
 This means that the mobility edge $E=E_m$, besides separating localized and delocalized eigenstates, provides also the boundary for the energy spectrum to remain real.
 The Lyapunov exponent analysis also provides the upper and lower boundaries of the energy spectrum in the delocalized phase, $E_{min}$ and $E_{max}$, shown in Fig.~1.
 In fact, since $E=2 t \Omega_{n_0}=2 t F(q=2 \pi \alpha n_0)$, from Eq.~(\ref{S2-7}) one has
 \begin{equation}
 E_{min}= {\rm min}_{-\pi \leq q < \pi} \left\{ 2t  F(q) \right\}=-\frac{2t}{t+1}
 \end{equation}
and
 \begin{equation}
 E_{max}= {\rm max}_{-\pi \leq q < \pi} \left\{ 2t  F(q) \right\}=\frac{2t}{t-1}.
 \end{equation}

\section{ Topological signature of mobility edges}
The appearance of mobility edges, separating extended and localized states, can be characterized by a topological invariant, given by a winding number that measures the times the spectral trajectory of the system encircle a given base energy $E_B$ when an additional phase in the potential is varied \cite{Gong,Longhi1,Zeng3}. Here we discuss in details how the winding numbers can be introduced for the two models discussed in main text.\\
{\em Model I.} As discussed in the main text, the existence of a mobility edge in Model I separates delocalized states with real energies and localized states with complex energies.
In particular, for a given value of $t=\exp(s)$, a mobility edge with energy $E_m=t(V-1)$ inside the energy spectrum, $E_{min}<E_m<E_{max}$, is found provided that
the potential amplitude $V$ is bounded as $V_1<V<V_2$, with (Fig.~1)
\begin{equation}
V_1= 1-\frac{2}{t+1} \; , \; V_2=1+\frac{2}{t-1}.
\end{equation}
To provide a topological characterization of the existence of the mobility edge, we consider Eq.~(1) on a ring lattice comprising $L$ sites with periodic boundary conditions and add an extra-phase term $\theta$ to the potential, i.e. we consider the eigenvalue equation
\begin{equation}
E \psi_n= t \sum_{n' \neq n}e^{-s\mid n-n'\mid}\psi_{n'}+V e^{i 2\pi\alpha n+i \theta }\psi_n \equiv H(\theta) \psi_n
\end{equation}
with Hamiltonian $H=H(\theta)$. For a given base energy $E_B$, we can introduce a winding number $w$ as follows~[2,3]
\begin{equation}
w(E_B)=\lim_{L \rightarrow \infty} \frac{1}{2 \pi i} \int_{0}^{2 \pi} d \theta \frac{\partial}{\partial \theta} {\rm log \; det} \left\{ H \left( \frac{\theta}{L} \right)-E_B \right\}
\label{wb}
\end{equation}

\begin{figure}
  \centering
  \includegraphics[width=0.45\textwidth]{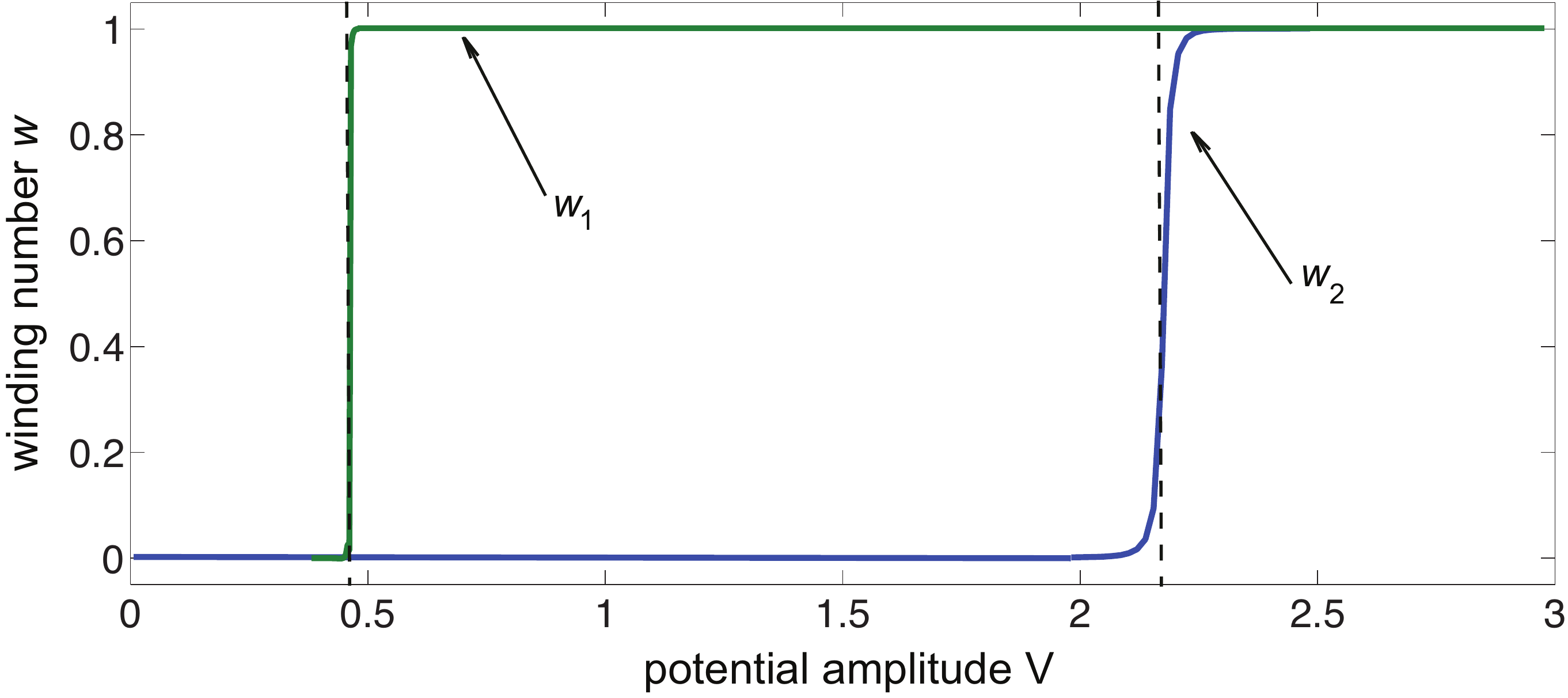}\\
  \caption{(Color online)  Numerically-computed behavior of the winding numbers $w_1$ and $w_2$ versus potential amplitude $V$ for Model I with $s=1$. The energy-dependent mobility edge is found between the two vertical dashed lines at $V=V_1$ and $V=V_2$, where the winding numbers $w_1$ and $w_2$ undergo an abrupt change from 0 to 1}
  \label{FigS4}
\end{figure}

which counts the number of times the complex spectral trajectory encircles the base energy point $E_B$
when the phase $\theta$ varies from zero to $2 \pi$. Clearly, we expect $w(E_B)$ to vanish when the energy spectrum is entirely real, but also when the energy spectrum can be partially complex but the base energy $E_B$ is smaller (larger) than the real part of any eigenvalue. Therefore, an appropriate topological characterization of the mobility edge requires to introduce two winding numbers $w_1=w(E_B=E_{min})$ and $w_2=w(E_B=E_{max})$, where $E_{min}$, $E_{max}$ are the lower and upper edges of the energy spectrum in the delocalized phase (Fig.~1). The numerically-computed behavior of $w_1$ and $w_2$ versus $V$, shown in Fig.~5, clearly indicates that the topological number $W=w_1 (1-w_2)$ is non-vanishing and equals one solely for $V_1<V<V_2$, i.e when there is an energy-dependent mobility edge.\\
{\em Model II.} For Model II, we consider the spectral problem for the Hamiltonian $H=H(\theta)$, which includes a phase shift $\theta$ in the potential, given by
\begin{eqnarray}
E \psi_n & = &  \psi_{n+1}+\psi_{n-1}+ \frac{V}{1-a e^{2 \pi i \alpha n+ i \theta}} \psi_n  \nonumber \\
& \equiv  & H(\theta) \psi_n.
\end{eqnarray}
For  a given base energy $E_B$, we can introduce a winding number $w(E_B)$ according to Eq.~(\ref{wb}). As noticed in the main text, for a given tuning parameter $a$ the mobility edge $E_m$ is independent of the potential strength $V$ and given by $E_m=a+1/a$. Owing to such a peculiar property, a single winding number can provide a topological signature of the mobility edge, which is obtained by assuming a base energy infinitesimally larger than $E_m$, i.e. $E_B=E_m^+$. As the potential strength $V$ is increased above zero, the winding number $w(E_B)$ is equal to one in the range $V_1<V<V_2$, where $V_1$ and $V_2$ are intersections of the mobility edge energy $E=E_m$ with the boundaries of the real part of the energy spectrum. This behavior is illustrated in Fig.~6, which shows the numerically-computed behavior of winding number $w$ versus $V$. Note that below $V_1$ (above $V_2$), where all eigenstates are extended (localized), the winding number vanishes.
\begin{figure}
  \centering
  \includegraphics[width=0.45\textwidth]{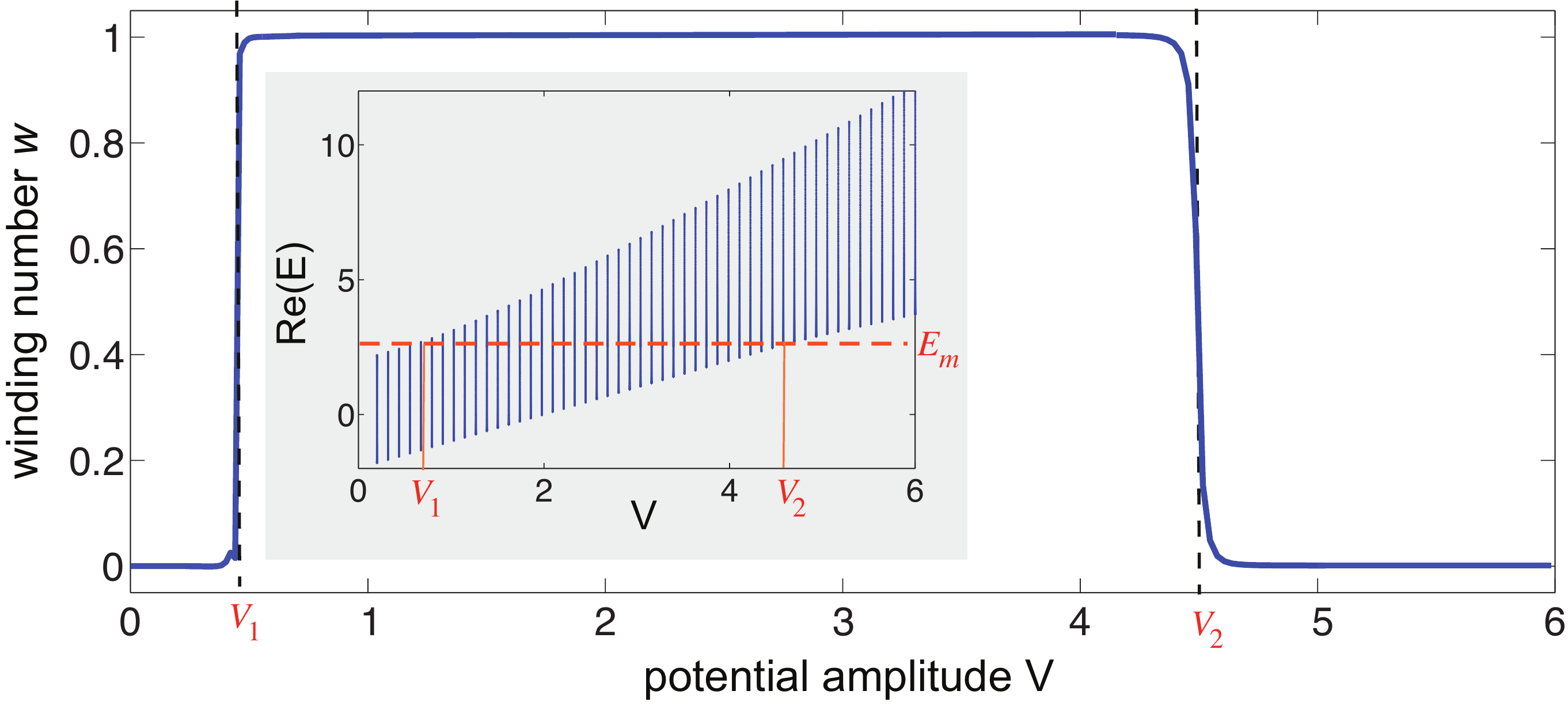}\\
  \caption{(Color online)  Numerically-computed behavior of the winding number $w$ versus potential amplitude $V$ for Model II with $a=0.5$, corresponding to a flat mobility edge $E_m=2.5$. The base energy used to compute the winding number is $E_B=2.51$. The inset shows the behavior of the real part of the energy spectrum versus $V$. The crossing point $V=V_1$ ($V=V_2$) corresponds to the potential amplitude below (above) which all eigenstates are extended (localized).}
  \label{FigS5}
\end{figure}

\section {Photonic implementation of Model II}
In this section we present the main model that describe the discrete-time quantum walk of optical pulses in the synthetic photonic mesh lattice realized by the fiber optical setup
described in Fig.~3. The analysis is rather standard and is an extension of derivations provided in previous works (see, for instance, \cite{RS7} which largely inspired our setup).  At each transit in the main loop, a pulse  entering into the interferometer  is splitted, at the output port, into three pulses with time delays $-\Delta t$, $0$ and $\Delta t$, where $\Delta t=\Delta L/c$ is the time delay introduced by the unbalanced arms in the interferometer. The successive pulse splitting emulates a discrete-time quantum walk, where the complex amplitude $a_n^{(m)}$ of pulse occupying the $n$-th time slot (discrete space distance) at the $m$-th round trip evolves according a linear map \cite{RS1,RS2,RS3,RS4,RS5,RS6,RS7,RS8}. The symmetric $3 \times 3$ fiber couplers between the three-arm interferometer and the main fiber loop is described by the scattering matrix \cite{RS9}
\begin{equation}
S^{(3)}= \frac{1}{\sqrt{3}} \left(
\begin{array}{ccc}
1 & \exp(i \theta) & \exp(i \theta) \\
\exp(i \theta) & 1 &  \exp(i \theta) \\
\exp(i \theta) & \exp(i \theta) & 1
\end{array}
\right) \label{3splitter}
\end{equation}
with $ \theta= -2 \pi /3$, while the $2 \times 2$ fiber coupler used to inject the initial pulse is described by the scattering matrix
\begin{equation}
S^{(2)}= \frac{1}{\sqrt{2}}\left(
\begin{array}{cc}
1 & i \\
i & 1
\end{array}
\right).  \label{2splitter}
\end{equation}
From Fig.~3 and using Eqs.~(\ref{3splitter}) and~(\ref{2splitter}), the following map can be readily obtained
\begin{equation}
a^{(m+1)}_n=\frac{\exp(G)}{3 \sqrt{2}} \left[ \exp(2 i \theta) \left( a_{n+1}^{(m)}+a_{n-1}^{(m)}   \right)  +U_n a^{(m)}_n  \right]
\label{mappa}
\end{equation}
where we have set
\begin{equation}
U_n=\exp (g-h^{(AOM)}_n-i h^{(EOM)}_n).
\end{equation}
\begin{figure}
  \centering
  \includegraphics[width=0.45\textwidth]{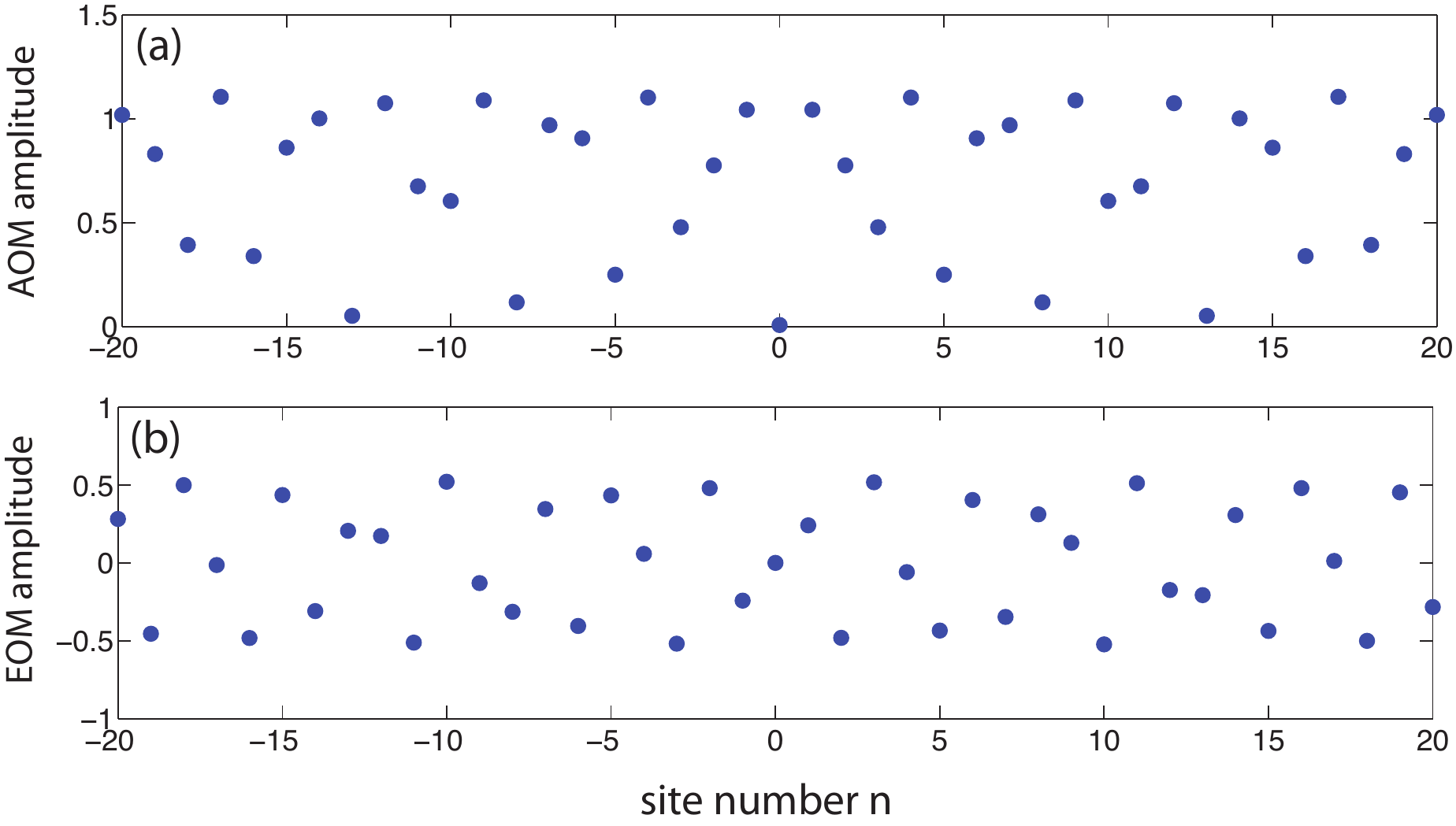}\\
  \caption{(Color online) Patterns of (a) AOM and (b) EOM signals ($h^{(AOM)}_n$ and $h^{(EOM)}_n$) that realize the complex potential $V_n$ of model II for parameter values $V=1$ and $a=0.5$ [Fig.~4(b) of the main text]. The gain $g$ is set to $g=1$. In (b), the constant bias $-2 \theta$ to $h_n^{(EOM)}$ has been omitted for the sake of simplicity. }
\end{figure}

In the above equations, $G$ and $g$ are the gain coefficients of the SOA in the main loop and in the central arm of the interferometer, respectively, whereas $h^{(AOM)}_n$
and $h^{(EOM)}_n$ are the amplitudes of AOM and EOM modulators, respectively, at the $n$-th time slot. To create a time-domain analog of a complex optical $U_n$, the EOM and AOM modulators  are
independently  driven with an waveform generator. The waveform is a specially designed
stepwise pulse pattern that enables to generate arbitrary phase and amplitude distributions along the fast coordinate (the time slot $n$) but constant along
the slow coordinate \cite{RS7}, i.e. periodic at time intervals $T=(L+L_0)/c$. The spectrum and localization properties of the eigenstates of the map Eq.~(\ref{mappa}) are obtained by taking the Ansatz
\begin{equation}
a_n^{(m)}= \mu^m \psi_n
\end{equation}
which yields the matrix spectral problem
\begin{equation}
E \psi_n= \psi_{n+1}+\psi_{n-1}+V_n \psi_n
\end{equation}
where we have set
\begin{equation}
V_n=U_n \exp(-2 i \theta)=\exp \left[  -2i \theta+g-h^{(AOM)}_n-ih^{(EOM)}_n \right]
\end{equation}
and $E=3 \sqrt{2} \mu  \exp(-G-2 i \theta)$, i.e.
\begin{equation}
\mu=\frac{E}{3 \sqrt{2}} \exp(G+2i \theta).
\end{equation}
Note that the gain $G$ supplied by the SOA in the main loop controls the growth/decay rate $|\mu |$ of the pulse train at successive transits, however it does not affect the shape of the complex potential $V_n$. The gain $G$ should be tuned to keep the system below the instability (lasing) threshold yet close to the threshold point in order to have enough signal at the detection output to monitor the pulse spreading dynamics in the lattice for several transits \cite{RS7}.\\
To reproduce the Model II, i.e. to realize the complex potential $V_n=V/[1-a \exp(2 \pi i \alpha n)]$ with $0<a<1$, the amplitudes $h^{(AOM)}_n$ and $h^{(EOM)}_n$ of the modulators should be set as follows
\begin{eqnarray}
h_n^{(EOM)} & = & -2 \theta+  {\rm atan} \frac{a \sin (2 \pi \alpha n)}{1-a \cos (2 \pi \alpha n)} \\
h_n^{(AOM)} & = &g + \log  \frac{\sqrt{1+a^2-2 a \cos (2 \pi \alpha n)}}{V}. \;\;\;\;\;\;
\end{eqnarray}
As an example, Fig.~7 shows the patterns of the EOM and AOM amplitudes corresponding to parameter values $V=1$, $a=0.5$ and $g=0.7$, i.e. to parameter values used in the simulations of Fig.~4(a).

\end{document}